\documentclass[aps,prl,amsmath,twocolumn,floats,floatfix,superscriptaddress,nofootinbib,showpacs]{revtex4-1}

\usepackage{amsfonts,amsmath,units,wasysym,epsfig,graphicx,verbatim,color,subfigure,graphicx}
\usepackage{amsmath}
\usepackage{amssymb}
\usepackage{amsfonts}
\usepackage{hyperref}
\usepackage{bm}
\usepackage{color}
\usepackage[normalem]{ulem}
\usepackage{natbib}
\usepackage{graphicx}
\usepackage{xcolor}

\def\be{\begin{equation}}
\def\ee{\end{equation}}
\def\bea{\begin{eqnarray}}
\def\eea{\end{eqnarray}}

\begin{document}
\bibliographystyle{unsrt}


\newcommand{\rhat}{\hat{r}}
\newcommand{\iotahat}{\hat{\iota}}
\newcommand{\phihat}{\hat{\phi}}
\newcommand{\h}{\mathfrak{h}}
\newcommand{\vek}[1]{\boldsymbol{#1}}
\newcommand{\IUCAA}{\affiliation{Inter-University Centre for Astronomy and Astrophysics, Post Bag 4, Ganeshkhind, Pune 411 007, India}}
\newcommand{\WSU}{\affiliation{Department of Physics \& Astronomy, Washington State University, 1245 Webster, Pullman, WA 99164-2814, U.S.A}}
\newcommand{\AEI}{\affiliation{Max Planck Institute for Gravitational Physics (Albert Einstein Institute), Hannover, Germany}} 

\title{News from horizons in binary black hole mergers}

\author{Vaishak Prasad}
\affiliation{Inter-University Centre for Astronomy and Astrophysics, Post Bag 4, Ganeshkhind, Pune 411 007, India}

\author{Anshu Gupta}
\affiliation{Inter-University Centre for Astronomy and Astrophysics, Post Bag 4, Ganeshkhind, Pune 411 007, India}

\author{Sukanta Bose}
\affiliation{Inter-University Centre for Astronomy and Astrophysics, Post Bag 4, Ganeshkhind, Pune 411 007, India}
\affiliation{Department of Physics and Astronomy, Washington State University, 1245 Webster, Pullman, WA 99164-2814, U.S.A}

\author{Badri Krishnan}
\affiliation{Max-Planck-Institut f\"ur Gravitationsphysik (Albert Einstein Institute), Callinstr. 38, 30167 Hannover, Germany}
\affiliation{Leibniz  Universit\"at  Hannover,  Welfengarten  1-A,  D-30167  Hannover,  Germany}

\author{Erik Schnetter}
\affiliation{Perimeter Institute for Theoretical Physics, Waterloo, Ontario, Canada}
\affiliation{Department of Physics \& Astronomy, University of Waterloo, Waterloo, Ontario, Canada}
\affiliation{Center for Computation \& Technology, Louisiana State University, Baton Rouge, Louisiana, USA}

\date{\today}

\begin{abstract}

  In a binary black hole merger, it is known that the
  inspiral portion of the waveform corresponds to two distinct horizons
  orbiting each other, and the merger and ringdown signals correspond 
  to the final horizon being formed and settling down to equilibrium.  
  However, we still lack a detailed understanding of the relation 
  between the horizon geometry in these three regimes and the observed 
  waveform. Here we show that the well known inspiral 
  chirp waveform has a clear counterpart on black hole horizons, 
  namely, the shear of the outgoing null rays at the horizon.  
  We demonstrate that the shear behaves very much like a compact binary coalescence waveform 
  with increasing frequency and amplitude. Furthermore, the parameters of the system estimated from 
  the horizon agree with those estimated from the waveform.  This implies
  that even though black hole horizons are causally disconnected from us, 
  assuming general relativity to be true, we can potentially infer some of their 
  detailed properties from gravitational wave observations. {\color{red}[This document has been assigned the LIGO Preprint number LIGO-P2000098.]} 

\end{abstract}

\maketitle

\noindent \emph{Introduction:} Starting with the first binary black
hole detection in 2015 \cite{Abbott:2016blz}, at least 10 binary black
hole mergers have been observed to date
\cite{LIGOScientific:2018mvr,TheLIGOScientific:2016pea,Nitz:2018imz,Nitz:2019hdf,Venumadhav:2019lyq,Zackay:2019tzo}.
For all of these detections, the parameters of the binary system,
including the masses and spins of the individual black holes, can be
inferred from the observed data \cite{TheLIGOScientific:2016wfe}. 
This inference relies crucially on gravitational waveform models
meant to represent, with sufficient accuracy, the gravitational 
wave emission from binary black hole mergers in general relativity
\cite{PhysRevLett.113.151101,PhysRevD.91.024043,PhysRevD.95.044028}.
Within general relativity, we also have detailed information
about properties of curved spacetime around a black hole merger from numerical simulations.
Since the first successful merger simulations
\cite{Pretorius:2005gq,Campanelli:2005dd,Baker:2005vv}, it is now
relatively straightforward to evolve black hole binaries through the
inspiral, merger and ringdown regimes, at least for moderate 
mass ratios.  Indeed, the waveform models mentioned above are all based 
on, and ultimately verified by, comparisons with these numerical 
simulations.  The wealth of information contained in the full numerically generated binary 
black hole spacetimes might plausibly have some imprints in the observed gravitational wave signal.

One might in fact be able to infer properties of spacetime regions
hidden behind the event horizon and causally disconnected from us.
The signal received at our observatories is generated by the
non-linearities and dynamics of the spacetime metric near and
around the black holes.  These non-linear and dynamical fields are
responsible for both the signal seen by us, and also the
properties of spacetime inside the event horizon \cite{Jaramillo:2011rf,Jaramillo:2011re,Jaramillo:2012rr,Gupta:2018znn}.  
The infalling flux of gravitational waves, representing tidal coupling, is of course part of the energy balance governing the dynamics of the binary system.  Thus, any modifications of the infalling flux will also affect the overall dynamics of the system.   
 Furthermore, in situations when the black holes are spinning sufficiently rapidly, 
phenomena like superradiance can play an important role (see e.g. \cite{Brito:2015oca}).   
In the regime when the effect of the companion can be treated 
as a perturbation, these tidal effects can be calculated 
analytically \cite{Hawking:1972hy,Hartle:1974gy,Hartle:1973zz,OSullivan:2015lni,OSullivan:2014ywd,PhysRevD.70.084044,Chatziioannou:2016kem,Chatziioannou:2012gq}.  Here we shall go beyond these calculations, close to the merger where linear perturbation theory is not sufficient. Note also that all of these perturbative calculations refer to the event horizon for which, as a matter of principle, no generally valid non-perturbative quasi-local flux formula can exist.  As in almost all such numerical studies, we always work with dynamical horizons which are expected to differ significantly from the event horizon near the merger, and do not suffer from the teleological properties of the event horizon.  Moreover, exact flux formulae, valid in full non-linear general relativity, are known for dynamical horizons (see e.g. \cite{Ashtekar:2004cn,Booth:2005qc}).  

Besides the effects on the orbital motion through tidal coupling, 
the infalling radiation must be very special for other reasons.  
The remnant black hole horizon is highly distorted on 
formation, and it loses its hair to reach its 
final equilibrium state represented by a Kerr black hole.  
However, the horizon, being a one-way membrane, cannot ``radiate away" 
its hair.  Instead, it approaches its very special final 
state by absorbing just the right amount 
of infalling radiation 
that
precisely cancels any hair that it might have when
it 
is initially formed \cite{Gupta:2018znn}.  Thus, the infalling radiation 
also determines the highly non-trivial issue of the final state within 
general relativity.  The important question then is this: Assuming that general 
relativity holds, are there strong correlations between the observed 
gravitational wave signal, and suitable fields on black hole horizons?  
If we have a sufficiently detailed understanding of these 
correlations, we might be able to understand fundamental properties of 
dynamical strong field gravity and black hole horizons 
from gravitational wave observations.

The presence of such correlations might be difficult to
discern in analytic or numerical studies.  The relevant spacetime
regions could not be more different: The spacetime we inhabit is very
close to flat and extremely well described by linearized general
relativity.  On the other hand, the region where black hole horizons 
live could have very high curvature (depending on the mass of
the black hole), and non-linearities of the Einstein equations need to
be taken into account.  It is thus not immediately obvious
precisely which fields should be correlated in a gauge invariant manner, and how mathematical results might be proved.  

Despite these potential pitfalls, several authors have previously found
evidence for correlations between quantities on horizons and in the
wave-zone
\cite{Jaramillo:2011rf,Jaramillo:2012rr,Rezzolla:2010df,Gupta:2018znn}.  These works have considered either the post-merger regime or subtleties regarding gravitational 
wave recoil.  Thus far none have considered what is one of the most well known features of binary merger waveforms, namely the inspiral chirp with increasing frequency and amplitude. The evolution of the frequency and amplitude have been calculated to high orders in various
post-Newtonian approaches, while accounting for a variety of physical effects such as
precession and eccentricity  (see, e.g., \cite{Blanchet2014}).   Moreover, these post-Newtonian calculations have been combined with numerical relativity merger signals to construct complete waveform models including the inspiral, merger and ringdown regimes as well  \cite{Ajith:2007kx,Ajith:2009bn,Santamaria:2010yb,Buonanno:1998gg,Bohe:2016gbl,Khan:2018fmp,Khan:2019kot}.  We shall use these complete waveform models to quantitatively compare gravitational wave signals with horizon fields.  

It has been shown previously
\cite{Gupta:2018znn} that for the post-merger signal, the
gravitational wave News (in essence the time derivative of the
gravitational wave strain), is correlated with the shear
on the black hole horizons. Here we will extend this study
to the inspiral regime, and show quantitatively that the chirp
signal is also extremely well correlated with the News.  
Remarkably, for reasons that we do not yet fully understand, we shall see
that very little effort is required to extract these correlations, 
and the gauge conditions employed in the simulations do not seem to play an important role.


\noindent \emph{Basic notions:} Our results deal with two
surfaces. The first is future null infinity $\mathcal{I}^+$, the end
point of future null-geodesics which escape to infinity 
\cite{Bondi:1962px,Penrose:1964ge}.  The second
is a dynamical horizon $\mathcal{H}$ \cite{Ashtekar:2004cn,Booth:2005qc} 
obtained by a time evolution of marginally trapped surfaces. These 
two surfaces might seem initially to be very different.  Future null
infinity $\mathcal{I}^+$ is an invariantly defined null surface 
where outgoing null geodesics end.
On the other hand, a dynamical horizon is located
inside the event horizon.  Nevertheless, both $\mathcal{I}^+$ and
$\mathcal{H}$ are one-way membranes and exact flux formulae hold for
both surfaces.  


For both cases, we consider spacelike 2-surfaces $S$ of spherical
topology, with an intrinsic Riemannian metric $q_{ab}$.  
$S$ will be either a cross section of $\mathcal{I}^+$ (approximated as a large coordinate sphere in the wave-zone enclosing the
source), or a section of $\mathcal{H}$, i.e. a marginally trapped surface.  In
either case, assuming that we can assign outgoing and ingoing
directions, we denote the outgoing future directed null vector normal
to $S$ by $\ell^a$, and the ingoing null normal as $n^a$; we will
require $\ell\cdot n = -1$.  Let $m$ be a complex null vector tangent to $S$
satisfying $m\cdot\bar{m}=1$ (the overbar denotes complex conjugation),
and $\ell\cdot m = n\cdot m = 0$.  

In the wave-zone, spacetime geometry is completely described by the 
Weyl tensor $C_{abcd}$.  In particular, outgoing transverse radiation 
is described by the Weyl tensor component \cite{Newman:1961qr}
\begin{equation}
 \Psi_4 = C_{abcd}n^a\bar{m}^bn^c\bar{m}^d\,.
\end{equation}
$\Psi_4$ can be expanded
in spin-weighted spherical harmonics ${}_{-2}Y_{\ell,m}$ of spin weight $-2$
\cite{Goldberg:1966uu}.  Let $\Psi_4^{(\ell,m)}$ be the mode component
with $\ell\geq 2$ and $-m\leq \ell \leq m$.  The $(\ell,m)$ component
of the News function $\mathcal{N}^{(\ell,m)}$ is defined as
\cite{Bondi:1962px}
\begin{equation}
  \mathcal{N}^{(\ell,m)}(u) = \int_{-\infty}^u\Psi_4^{(\ell,m)}\,du\,.
\end{equation}
The outgoing energy flux is related to the integral of $|\mathcal{N}|^2$ over 
all angles.  In a numerical spacetime it is in
principle possible to extract $\Psi_4$ going out all the way to
$\mathcal{I}^+$ \cite{Babiuc:2008qy}, and this is what should be done
to reduce systematic errors.  We shall follow the common approach of
calculating $\Psi_4$ on a sphere at a finite radial coordinate $r$ and
the integral in the previous equation is over time instead of the
retarded time coordinate $u$. The lower limit in the integral is not
$-\infty$ but the earliest time available in the simulation.  The News
function is then a function of time at a fixed value of $r$, starting
from the earliest time available in the simulation. A further time
integration of $\mathcal{N}$ yields the gravitational wave strain.  

Turning now to the black hole, the basic object here is a marginally
outer trapped surface (MOTS), again denoted $S$.  This is a closed spacelike
2-surface with vanishing outgoing expansion $\Theta_{(\ell)}$: 
\begin{equation}
  \Theta_{(\ell)} = q^{ab}\nabla_a\ell_b = 0\,.
\end{equation}
The shear of $\ell^a$ is defined as
\begin{equation}
  \sigma = m^am^b\nabla_a\ell_b\,.
\end{equation}
Both $\mathcal{N}$ and $\bar{\sigma}$ have the same behavior under
spin rotations $m\rightarrow me^{i\psi}$, i.e. they have the same spin
weight. Also, similar to the news function and the Bondi mass-loss
formula, $|\sigma^2|$ appears in the energy flux falling into the
black hole \cite{Ashtekar:2002ag,Ashtekar:2003hk}, though in this case
the flux also contains other contributions. It is shown
in \cite{Booth:2003ji} that for the case of a slowly evolving horizon,
which is what we are dealing with in the inspiral
phase, $ |\sigma|^2$ is the dominant part of the flux. Thus, as suggested in
\cite{Jaramillo:2011rf}, we will compare the shear at the horizon with
the News.    

\noindent \emph{The numerical simulations:} Our numerical simulations are 
performed using the publicly available Einstein Toolkit framework 
\cite{Loffler:2011ay, EinsteinToolkit:web}. The initial data is generated 
based on the puncture approach \cite{PhysRevLett.78.3606,Ansorg:2004ds}, 
which has been evolved through  BSSNOK formulation
\cite{Alcubierre:2000xu, Alcubierre:2002kk, Brown:2008sb} using the
$1+\log$ slicing and $\Gamma$-driver shift conditions.  Gravitational 
waveforms are extracted \cite{Baker:2002qf} on coordinate spheres at 
various radii between $100 M$ to $500 M$.  
The computational grid set-up is based on the multipatch approach using 
Llama \cite{PhysRevD.83.044045} and Carpet modules, along with adaptive 
mesh refinement (AMR).  The various horizons (or more precisely, 
marginally outer trapped surfaces) are located using the method 
described in \cite{Thornburg:1995cp, Thornburg:2003sf}. 
Quasi-local physical quantities are computed on the horizons following 
\cite{Dreyer:2002mx, Schnetter:2006yt}.

We consider non-spinning binary black hole systems with varying mass-ratio 
$q=M_2/M_1$, where $M_{1,2}$ are the component masses (with $M_1\geq M_2$). 
We use the GW150914 parameter file available from \cite{wardell_barry_2016_155394} as
our template. For each of the simulations, as input parameters we provide initial
separation between the two punctures $D$, mass ratio $q$ and the radial and azimuthal linear momenta $p_r$, $p_{\phi}$ respectively, while keeping the total mass $M=M_1+M_2=1$. Parameters are listed in table \ref{tab:ic_qc0}. We then compute the corresponding initial locations, the $x$, $y$, $z$ components of linear momentum for both black holes, and grid refinement levels, etc., before generating the initial data and evolving it. We chose 6 non-spinning cases ranging between $q=1.0$ to $0.25$, based on the initial parameters listed in \cite{Healy:2014yta,PhysRevD.95.024037}.  Our simulations match very well with the catalog simulations \cite{RITcatalog:web}, having merger time discrepancies less than a few percent.

\begin{table}
\begin{tabular}{|p{2cm}|p{2cm}|p{2cm}|p{2cm}|}
 \hline
 $q $ & $D/M$ & $p_r/M$& $p_{\phi}/M$\\
 \hline
1.0&  9.5332&0.0 &0.099322\\
0.85 &12.0 &-0.000529 &0.08448\\
0.75 &11.0 &-0.000686 &0.08828\\
0.667 &11.75 & -0.000529 &0.08281\\
0.5 &11.0& -0.000572&0.0802\\
0.25 &11.0&-0.000308 &0.05794\\
 \hline
\end{tabular}
\caption{Initial parameters for non-spinning binary black holes with quasi-circular orbits. $q=M_2/M_1$ is mass ratio, $D$ is the initial separation between the two holes, $p_r$ and $p_{\phi}$ are radial and azimuthal linear momenta respectively.}
\label{tab:ic_qc0}

\end{table}



\noindent \emph{Results:} We begin by looking at the complex shears,
$\sigma_1$ and $\sigma_2$, of the outgoing null normal $\ell^a$ at the
two individual horizons for a particular configuration, namely
$q=0.25$.   We write the shear as $\sigma = \sigma_+ + i\sigma_\times$.  As in \cite{Gupta:2018znn}, we introduce coordinates $(\theta,\phi)$ on the
horizons with the $z$-axes perpendicular to the orbital plane. Just like the waveform the angular distribution is mostly quadrupolar, i.e. $\sigma\propto {}_{-2}Y_{2,2}(\theta,\phi)$.  It will then be sufficient for our purposes to just look at the values of $\sigma_{1,2}$
on the north poles of the two horizons. This will not suffice for precessing spins or when higher modes become more important.  In these more complicated cases the approach suggested in \cite{Ashtekar:2013qta} can be followed.  

\begin{figure}
  \includegraphics[width=\columnwidth]{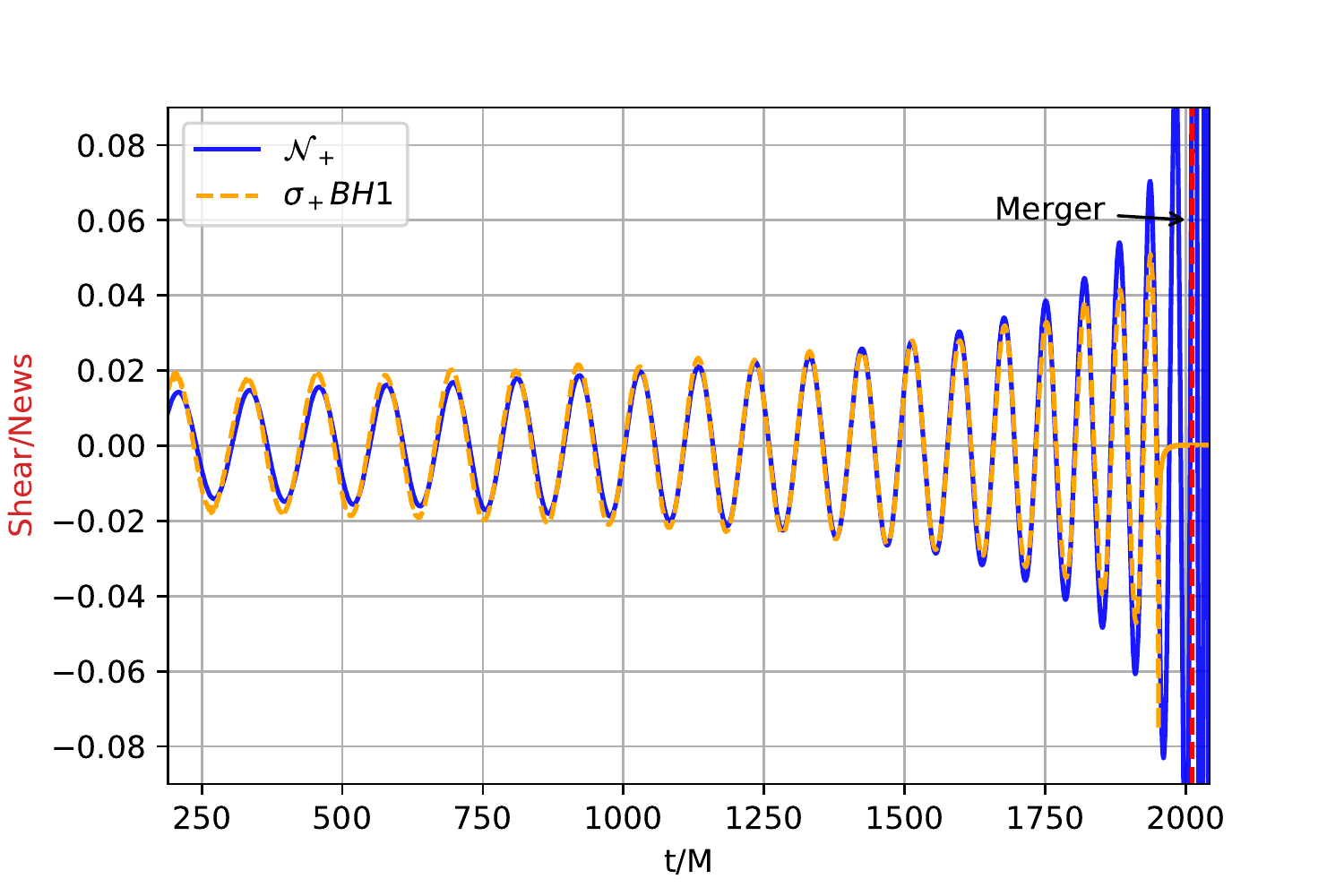}
  \includegraphics[width=\columnwidth]{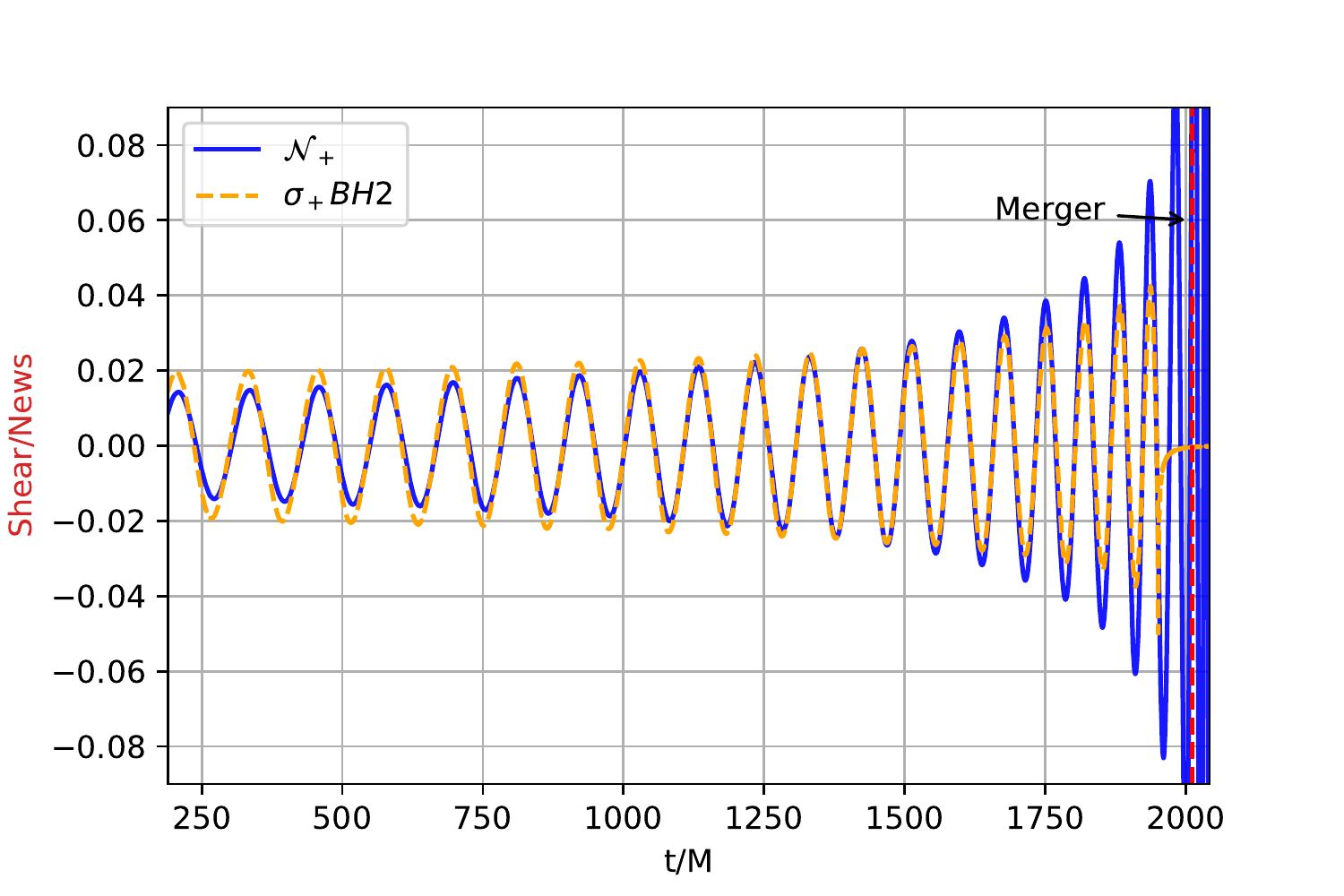}
  \caption{The real part of the shear, $\sigma_+(t)$, for the two black holes ($BH1$ and $BH2$) for the $q=0.25$ configuration, compared with the real part of the News function $\mathcal{N}_+^{(2,2)}$. The functions $\sigma_+$ and $\mathcal{N}_+^{(2,2)}$ have been suitably aligned and their amplitudes scaled so that they have unit norm over the time interval where the shear is defined. Similar results hold for the $\times$ polarization and the other mass-ratios.}
  \label{fig:shear-news}
\end{figure}

Figure~\ref{fig:shear-news} shows the real parts $\sigma_+(t)$ for the two black holes and for the $q=0.25$ configuration.  The plot also shows the News function for the corresponding mode and polarizations, i.e. $\mathcal{N}_+^{(2,2)}$.
The plot continues as long as the individual horizons can be
found reliably.  The time when the common horizon is formed is
indicated by a vertical line.  The figure shows that the
shear has very similar properties as the well known ``chirping''
gravitational waveform: it has increasing frequency and amplitude.  We
note also that since the shear is non-vanishing, it follows that the horizon is
not isolated and its area is increasing. However, the area
is not increasing rapidly and this area increase is not 
measured reliably in our simulations.  See 
\cite{PhysRevLett.123.171102,PhysRevD.100.084044} for a more 
accurate study of the area increase in a black hole merger.  

Does the qualitative agreement of the shear with gravitational wave
signals shown in Fig.~\ref{fig:shear-news} hold quantitatively?  To answer this question, we treat the
shear as a bonafide gravitational waveform and attempt to estimate its
parameters.  
The chirp mass, $\mathcal{M} = Mq^{3/5}/(1+q)^{6/5}$, determines the frequency
evolution of the signal at leading order.  For any given simulation,
we have then three possible gravitational wave signals: the waveform
$h(t)$ extracted in the wave-zone over a large sphere, and the 
shears $\sigma_{1,2}(t)$ calculated at the
individual horizons.  
Here $t$ is the coordinate time used in the
numerical evolution.  For all three of these time series, we estimate 
$\mathcal{M}$ and $q$ using a well tested model for binary
mergers known as IMRPhenomPV2
\cite{PhysRevD.93.044006,PhysRevD.93.044007}.  This waveform model is
a development of the so-called phenomenological binary merger models
\cite{Ajith:2007kx,Ajith:2009bn,Santamaria:2010yb} and it includes, in
principle, precession due to the misalignment of the individual spins
with the orbital angular momentum (though this is not relevant here). 
Other waveforms could also be used \cite{Buonanno:1998gg,Bohe:2016gbl,Khan:2018fmp,Khan:2019kot} 
but we do not expect any significant differences for our purposes.  

We obtain three estimates of $(q,\mathcal{M})$ using, in-turn, the
waveform
$h_{+,\times}$ extracted in the wave-zone (on an extraction
sphere of radius $R_E= 100M$), and the shears $\sigma_{1,2}$ at the two
horizons. The waveform $h_{+,\times}$ is matched with the
model waveform itself, while $\sigma_{1,2}$ are matched with the News, i.e., the
time derivative of the model waveform. In each case we use a
sufficiently fine grid in $(q,\mathcal{M})$ and minimize a standard least-squares
figure-of-merit over the relative time-shift and initial phase (which will henceforth be referred to as alignment) as well as the mass parameters.  Table~\ref{tab:bestfit} shows the best fit values of
$\mathcal{M}$ and $q$ for the real part of the shears.  The chirp mass $\mathcal{M}$ is very well
measured, with typical errors of $\sim 0.5\%$ for the strain and 
$\sim  1-3 \%$ for the shears.  The uncertainties in the mass-ratio 
are much larger (as expected), with errors of $\sim 9\%$ for the 
strain and $\sim 17-24\%$ for the shears.   We have chosen to 
present our results in terms of $(\mathcal{M},q)$ as independent 
parameters, though we could have used the total mass $M$ as well. 
It is easy to check that the best fit values of $M$ turn out to be 
very close to unity as they should. 

Motivated by this excellent agreement, we postulate that given the News, it should be possible to predict the horizon shears.  First comparing the amplitudes and the phases of $\sigma_1$ and $\sigma_2$ (after aligning them with the News), we find that to a very good approximation the phase difference between them is very small, and 
\begin{equation}
    \frac{M_2\sigma_2}{M_1\sigma_1} \approx q^{-0.7}\,.
\end{equation}
We multiply the shears by the respective masses to make them dimensionless.  
Similalrly, comparing the News and one of the shears, say $\sigma_1$, we find 
that the phase difference is again small and their amplitudes are related as follows:
\begin{equation}
    R_E\left|\mathcal{N}\right| \approx 0.5(1+q)M_1\left|\sigma_1\right|\,.
\end{equation}
Here $R_E=100M$ is the extraction radius.
With these relations, given the observed gravitational wave strain, one can estimate the amplitude and frequency content of the horizon shears for binaries consisting of non-spinning black holes.

\begin{table}{}
\begin{center}
\begin{tabular}{|p{1cm}|c|p{1cm}|p{1cm}|p{1cm}|p{1cm}|p{1cm}|}
\hline
$\widehat{q}$ & $q$ & $\mathcal{M}$ & $q_1$ & $\mathcal{M}_1$ & $q_2$ & $\mathcal{M}_2$ \\
\hline
$1.0$ & $1.0$ & $0.432$ & $0.800$& $0.439$ & $1.200$& $0.439$ \\ 
$0.85$ & $1.0$ & $0.433$& $0.770$& $0.434$ & $0.770$& $0.434$ \\ 
$0.75$ & $0.78$ & $0.428$& $0.630$& $0.426$ & $0.870$& $0.426$ \\ 
$0.67$ & $0.779$ &$0.427$& $0.867$& $0.433$ & $0.747$& $0.429$ \\ 
$0.50$ & $0.498$ & $0.403$& $0.580$& $0.410$ & $0.580$& $0.410$ \\ 
$0.25$ & $0.222$ &$0.328$& $0.330$& $0.345$ & $0.250$& $0.333$ \\ 
\hline
\end{tabular}
\caption{Best fit values of the mass ratio and chirp mass for
  i) the waveform extracted in the wave-zone (denoted $q$ and
  $\mathcal{M}$), ii) the shear of the first black hole ($q_1$ and
  $\mathcal{M}_1$), and iii) the shear of the second black hole ($q_2$
  and $\mathcal{M}_2$).  The mass ratio is nominally $\widehat{q}$ for
  the puncture initial data.  }
\label{tab:bestfit}
\end{center}
\end{table}

\noindent \emph{Conclusions:} We have shown quantitatively that in a
black hole merger, the shear of the horizons
behaves just like gravitational wave signals seen
in the wave-zone.  This adds an important ingredient to the idea that
there are strong correlations between gravitational wave signals seen
by gravitational wave detectors and suitable fields in the strong
field dynamical region near the black holes. 

Future work will extend this study in many directions, e.g. allowing spinning black holes, precession 
effects, and possibly super-radiance in the non-linear merger regime. 
The horizon shears are related to the variation of the horizon multipole 
moments.  One might therefore be able to relate the radiative multipole
moments to the horizon moments.  On dynamical horizons, various balance
laws are known relating the change in the horizon multipole moments to 
fluxes across the horizon \cite{Ashtekar:2013qta}.  Combining this with 
the idea of slowly evolving horizons \cite{Booth:2003ji,Booth:2006bn} might provide 
an interesting route to relate properties of waveforms with horizons~\cite{Datta:2019euh,Datta:2019epe} and 
to perhaps build better waveform models.  Note that in Fig.~\ref{fig:shear-news} the difference between the shear and the News becomes larger near the merger where the additional terms in the horizon flux law start to matter.  It will be important to compare the full flux at the horizon with the News.  

A deeper mathematical understanding 
of these observed correlations is still lacking. In particular, it is 
important to identify the precise spacetime region and the non-linearities 
that generate the gravitational waves seen at the horizons and in the 
wave-zone.  Elucidating the precise relationship of our results with the 
perturbative calculations of tidal coupling \cite{Hawking:1972hy,Hartle:1974gy,Hartle:1973zz,OSullivan:2015lni,OSullivan:2014ywd} 
is of great interest as well.  In particular we mention the work by O'Sullivan \& Hughes 
\cite{OSullivan:2015lni,OSullivan:2014ywd} that studies high mass-ratio 
systems perturbatively, and especially the effect on the geometry of 
the event horizon. They find a strong correlation between the shear of 
the horizon with the particle orbit (and thus with the observed waveform) 
which is broadly consistent with our results.


\noindent \emph{Acknowledgments:} We are grateful to Abhay Ashtekar, Ivan Booth 
and Jose-Luis Jaramillo for valuable discussions.
Research at Perimeter Institute is supported in part by the Government of Canada through the Department of Innovation, Science and Economic Development Canada and by the Province of Ontario through the Ministry of Colleges and Universities.
The numerical simulations 
were performed on the high performance supercomputer \emph{Perseus} at IUCAA.

\section{References}
\bibliography{references.bib}

\begin{thebibliography}{10}

\bibitem{Abbott:2016blz}
B.~P. Abbott et~al.
\newblock {Observation of Gravitational Waves from a Binary Black Hole Merger}.
\newblock {\em Phys. Rev. Lett.}, 116(6):061102, 2016.

\bibitem{LIGOScientific:2018mvr}
B.~P. Abbott et~al.
\newblock {GWTC-1: A Gravitational-Wave Transient Catalog of Compact Binary
  Mergers Observed by LIGO and Virgo during the First and Second Observing
  Runs}.
\newblock {\em Phys. Rev.}, X9(3):031040, 2019.

\bibitem{TheLIGOScientific:2016pea}
B.~P. Abbott et~al.
\newblock {Binary Black Hole Mergers in the first Advanced LIGO Observing Run}.
\newblock {\em Phys. Rev.}, X6(4):041015, 2016.
\newblock [erratum: Phys. Rev.X8,no.3,039903(2018)].

\bibitem{Nitz:2018imz}
Alexander~H. Nitz, Collin Capano, Alex~B. Nielsen, Steven Reyes, Rebecca White,
  Duncan~A. Brown, and Badri Krishnan.
\newblock {1-OGC: The first open gravitational-wave catalog of binary mergers
  from analysis of public Advanced LIGO data}.
\newblock {\em Astrophys. J.}, 872(2):195, 2019.

\bibitem{Nitz:2019hdf}
Alexander~H. Nitz, Thomas Dent, Gareth~S. Davies, Sumit Kumar, Collin~D.
  Capano, Ian Harry, Simone Mozzon, Laura Nuttall, Andrew Lundgren, and Márton
  Tápai.
\newblock {2-OGC: Open Gravitational-wave Catalog of binary mergers from
  analysis of public Advanced LIGO and Virgo data}.
\newblock 2019.

\bibitem{Venumadhav:2019lyq}
Tejaswi Venumadhav, Barak Zackay, Javier Roulet, Liang Dai, and Matias
  Zaldarriaga.
\newblock {New Binary Black Hole Mergers in the Second Observing Run of
  Advanced LIGO and Advanced Virgo}.
\newblock 2019.

\bibitem{Zackay:2019tzo}
Barak Zackay, Tejaswi Venumadhav, Liang Dai, Javier Roulet, and Matias
  Zaldarriaga.
\newblock {A Highly Spinning and Aligned Binary Black Hole Merger in the
  Advanced LIGO First Observing Run}.
\newblock 2019.

\bibitem{TheLIGOScientific:2016wfe}
B.~P. Abbott et~al.
\newblock {Properties of the Binary Black Hole Merger GW150914}.
\newblock {\em Phys. Rev. Lett.}, 116(24):241102, 2016.

\bibitem{PhysRevLett.113.151101}
Mark Hannam, Patricia Schmidt, Alejandro Boh\'e, Le\"{\i}la Haegel, Sascha
  Husa, Frank Ohme, Geraint Pratten, and Michael P\"urrer.
\newblock Simple model of complete precessing black-hole-binary gravitational
  waveforms.
\newblock {\em Phys. Rev. Lett.}, 113:151101, Oct 2014.

\bibitem{PhysRevD.91.024043}
Patricia Schmidt, Frank Ohme, and Mark Hannam.
\newblock Towards models of gravitational waveforms from generic binaries: Ii.
  modelling precession effects with a single effective precession parameter.
\newblock {\em Phys. Rev. D}, 91:024043, Jan 2015.

\bibitem{PhysRevD.95.044028}
Alejandro Boh\'e, Lijing Shao, Andrea Taracchini, Alessandra Buonanno,
  Stanislav Babak, Ian~W. Harry, Ian Hinder, Serguei Ossokine, Michael
  P\"urrer, Vivien Raymond, Tony Chu, Heather Fong, Prayush Kumar, Harald~P.
  Pfeiffer, Michael Boyle, Daniel~A. Hemberger, Lawrence~E. Kidder, Geoffrey
  Lovelace, Mark~A. Scheel, and B\'ela Szil\'agyi.
\newblock Improved effective-one-body model of spinning, nonprecessing binary
  black holes for the era of gravitational-wave astrophysics with advanced
  detectors.
\newblock {\em Phys. Rev. D}, 95:044028, Feb 2017.

\bibitem{Pretorius:2005gq}
Frans Pretorius.
\newblock {Evolution of Binary Black Hole Spacetimes}.
\newblock {\em Phys. Rev. Lett.}, 95:121101, 2005.

\bibitem{Campanelli:2005dd}
Manuela Campanelli, C.~O. Lousto, P.~Marronetti, and Y.~Zlochower.
\newblock {Accurate evolutions of orbiting black-hole binaries without
  excision}.
\newblock {\em Phys. Rev. Lett.}, 96:111101, 2006.

\bibitem{Baker:2005vv}
John~G. Baker, Joan Centrella, Dae-Il Choi, Michael Koppitz, and James van
  Meter.
\newblock {Gravitational wave extraction from an inspiraling configuration of
  merging black holes}.
\newblock {\em Phys. Rev. Lett.}, 96:111102, 2006.

\bibitem{Jaramillo:2011rf}
Jose~Luis Jaramillo, Rodrigo~P. Macedo, Philipp M{\"o}sta, and Luciano
  Rezzolla.
\newblock {Black-hole horizons as probes of black-hole dynamics II: geometrical
  insights}.
\newblock {\em Phys. Rev.}, D85:084031, 2012.

\bibitem{Jaramillo:2011re}
Jose~Luis Jaramillo, Rodrigo~Panosso Macedo, Philipp M{\"o}sta, and Luciano
  Rezzolla.
\newblock {Black-hole horizons as probes of black-hole dynamics I: post-merger
  recoil in head-on collisions}.
\newblock {\em Phys.Rev.}, D85:084030, 2012.

\bibitem{Jaramillo:2012rr}
J.~L. Jaramillo, R.~P. Macedo, P.~M{\"o}sta, and L.~Rezzolla.
\newblock {Towards a cross-correlation approach to strong-field dynamics in
  Black Hole spacetimes}.
\newblock {\em AIP Conf. Proc.}, 1458:158--173, 2011.

\bibitem{Gupta:2018znn}
Anshu Gupta, Badri Krishnan, Alex Nielsen, and Erik Schnetter.
\newblock {Dynamics of marginally trapped surfaces in a binary black hole
  merger: Growth and approach to equilibrium}.
\newblock {\em Phys. Rev.}, D97(8):084028, 2018.

\bibitem{Brito:2015oca}
Richard Brito, Vitor Cardoso, and Paolo Pani.
\newblock {Superradiance}.
\newblock {\em Lect. Notes Phys.}, 906:pp.1--237, 2015.

\bibitem{Hawking:1972hy}
S.~W. Hawking and J.~B. Hartle.
\newblock {Energy and angular momentum flow into a black hole}.
\newblock {\em Commun. Math. Phys.}, 27:283--290, 1972.

\bibitem{Hartle:1974gy}
James~B. Hartle.
\newblock {Tidal shapes and shifts on rotating black holes}.
\newblock {\em Phys. Rev.}, D9:2749--2759, 1974.

\bibitem{Hartle:1973zz}
James~B. Hartle.
\newblock {Tidal Friction in Slowly Rotating Black Holes}.
\newblock {\em Phys. Rev.}, D8:1010--1024, 1973.

\bibitem{OSullivan:2015lni}
Stephen O'Sullivan and Scott~A. Hughes.
\newblock {Strong-field tidal distortions of rotating black holes: II. Horizon
  dynamics from eccentric and inclined orbits}.
\newblock {\em Phys. Rev.}, D94(4):044057, 2016.

\bibitem{OSullivan:2014ywd}
Stephen O'Sullivan and Scott~A. Hughes.
\newblock {Strong-field tidal distortions of rotating black holes: Formalism
  and results for circular, equatorial orbits}.
\newblock {\em Phys. Rev.}, D90(12):124039, 2014.
\newblock [Erratum: Phys. Rev.D91,no.10,109901(2015)].

\bibitem{PhysRevD.70.084044}
Eric Poisson.
\newblock Absorption of mass and angular momentum by a black hole: Time-domain
  formalisms for gravitational perturbations, and the small-hole or slow-motion
  approximation.
\newblock {\em Phys. Rev. D}, 70:084044, Oct 2004.

\bibitem{Chatziioannou:2016kem}
Katerina Chatziioannou, Eric Poisson, and Nicolas Yunes.
\newblock {Improved next-to-leading order tidal heating and torquing of a Kerr
  black hole}.
\newblock {\em Phys. Rev.}, D94(8):084043, 2016.

\bibitem{Chatziioannou:2012gq}
Katerina Chatziioannou, Eric Poisson, and Nicolas Yunes.
\newblock {Tidal heating and torquing of a Kerr black hole to next-to-leading
  order in the tidal coupling}.
\newblock {\em Phys. Rev.}, D87(4):044022, 2013.

\bibitem{Ashtekar:2004cn}
Abhay Ashtekar and Badri Krishnan.
\newblock {Isolated and dynamical horizons and their applications}.
\newblock {\em Living Rev. Rel.}, 7:10, 2004.

\bibitem{Booth:2005qc}
Ivan Booth.
\newblock {Black hole boundaries}.
\newblock {\em Can. J. Phys.}, 83:1073--1099, 2005.

\bibitem{Rezzolla:2010df}
Luciano Rezzolla, Rodrigo~P. Macedo, and Jose~Luis Jaramillo.
\newblock {Understanding the 'anti-kick' in the merger of binary black holes}.
\newblock {\em Phys. Rev. Lett.}, 104:221101, 2010.

\bibitem{Blanchet2014}
Luc Blanchet.
\newblock Gravitational radiation from post-newtonian sources and inspiralling
  compact binaries.
\newblock {\em Living Reviews in Relativity}, 17(1):2, Feb 2014.

\bibitem{Ajith:2007kx}
P.~Ajith et~al.
\newblock {A Template bank for gravitational waveforms from coalescing binary
  black holes. I. Non-spinning binaries}.
\newblock {\em Phys. Rev.}, D77:104017, 2008.
\newblock [Erratum: Phys. Rev.D79,129901(2009)].

\bibitem{Ajith:2009bn}
P.~Ajith et~al.
\newblock {Inspiral-merger-ringdown waveforms for black-hole binaries with
  non-precessing spins}.
\newblock {\em Phys. Rev. Lett.}, 106:241101, 2011.

\bibitem{Santamaria:2010yb}
L.~Santamaria et~al.
\newblock {Matching post-Newtonian and numerical relativity waveforms:
  systematic errors and a new phenomenological model for non-precessing black
  hole binaries}.
\newblock {\em Phys. Rev.}, D82:064016, 2010.

\bibitem{Buonanno:1998gg}
A.~Buonanno and T.~Damour.
\newblock {Effective one-body approach to general relativistic two-body
  dynamics}.
\newblock {\em Phys. Rev.}, D59:084006, 1999.

\bibitem{Bohe:2016gbl}
Alejandro Bohé et~al.
\newblock {Improved effective-one-body model of spinning, nonprecessing binary
  black holes for the era of gravitational-wave astrophysics with advanced
  detectors}.
\newblock {\em Phys. Rev.}, D95(4):044028, 2017.

\bibitem{Khan:2018fmp}
Sebastian Khan, Katerina Chatziioannou, Mark Hannam, and Frank Ohme.
\newblock {Phenomenological model for the gravitational-wave signal from
  precessing binary black holes with two-spin effects}.
\newblock {\em Phys. Rev.}, D100(2):024059, 2019.

\bibitem{Khan:2019kot}
Sebastian Khan, Frank Ohme, Katerina Chatziioannou, and Mark Hannam.
\newblock {Including higher order multipoles in gravitational-wave models for
  precessing binary black holes}.
\newblock {\em Phys. Rev.}, D101(2):024056, 2020.

\bibitem{Bondi:1962px}
H.~Bondi, M.~G.~J. van~der Burg, and A.~W.~K. Metzner.
\newblock {Gravitational waves in general relativity. 7. Waves from
  axisymmetric isolated systems}.
\newblock {\em Proc. Roy. Soc. Lond.}, A269:21--52, 1962.

\bibitem{Penrose:1964ge}
R.~Penrose.
\newblock {Conformal treatment of infinity}.
\newblock {\em Gen. Rel. Grav.}, 43:901--922, 2011.
\newblock [,565(1964)].

\bibitem{Newman:1961qr}
Ezra Newman and Roger Penrose.
\newblock {An Approach to gravitational radiation by a method of spin
  coefficients}.
\newblock {\em J. Math. Phys.}, 3:566--578, 1962.

\bibitem{Goldberg:1966uu}
J.~N. Goldberg, A.~J. MacFarlane, E.~T. Newman, F.~Rohrlich, and E.~C.~G.
  Sudarshan.
\newblock {Spin s spherical harmonics and edth}.
\newblock {\em J. Math. Phys.}, 8:2155, 1967.

\bibitem{Babiuc:2008qy}
M.~C. Babiuc, N.~T. Bishop, B.~Szilagyi, and J.~Winicour.
\newblock {Strategies for the Characteristic Extraction of Gravitational
  Waveforms}.
\newblock {\em Phys. Rev.}, D79:084011, 2009.

\bibitem{Ashtekar:2002ag}
Abhay Ashtekar and Badri Krishnan.
\newblock {Dynamical horizons: Energy, angular momentum, fluxes and balance
  laws}.
\newblock {\em Phys. Rev. Lett.}, 89:261101, 2002.

\bibitem{Ashtekar:2003hk}
Abhay Ashtekar and Badri Krishnan.
\newblock {Dynamical horizons and their properties}.
\newblock {\em Phys. Rev.}, D68:104030, 2003.

\bibitem{Booth:2003ji}
Ivan Booth and Stephen Fairhurst.
\newblock {The first law for slowly evolving horizons}.
\newblock {\em Phys. Rev. Lett.}, 92:011102, 2004.

\bibitem{Loffler:2011ay}
Frank L{\"{o}}ffler, Joshua Faber, Eloisa Bentivegna, Tanja Bode, Peter Diener,
  Roland Haas, Ian Hinder, Bruno~C. Mundim, Christian~D. Ott, Erik Schnetter,
  Gabrielle Allen, Manuela Campanelli, and Pablo Laguna.
\newblock {{T}he {E}instein {T}oolkit: {A} {C}ommunity {C}omputational
  {I}nfrastructure for {R}elativistic {A}strophysics}.
\newblock {\em Class. Quantum Grav.}, 29(11):115001, 2012.

\bibitem{EinsteinToolkit:web}
{Einstein Toolkit}: Open software for relativistic astrophysics.
\newblock \url{http://einsteintoolkit.org/}.

\bibitem{PhysRevLett.78.3606}
Steven Brandt and Bernd Br\"ugmann.
\newblock A simple construction of initial data for multiple black holes.
\newblock {\em Phys. Rev. Lett.}, 78:3606--3609, May 1997.

\bibitem{Ansorg:2004ds}
Marcus Ansorg, Bernd Br{\"u}gmann, and Wolfgang Tichy.
\newblock A single-domain spectral method for black hole puncture data.
\newblock {\em Phys. Rev. D}, 70:064011, 2004.

\bibitem{Alcubierre:2000xu}
Miguel Alcubierre, Gabrielle Allen, Bernd Br{\"u}gmann, Thomas Dramlitsch,
  Jose~A. Font, Philippos Papadopoulos, Edward Seidel, Nikolaos Stergioulas,
  Wai-Mo Suen, and Ryoji Takahashi.
\newblock {Towards a stable numerical evolution of strongly gravitating systems
  in general relativity: The Conformal treatments}.
\newblock {\em Phys. Rev.}, D62:044034, 2000.

\bibitem{Alcubierre:2002kk}
Miguel Alcubierre, Bernd Br{\"u}gmann, Peter Diener, Michael Koppitz, Denis
  Pollney, Edward Seidel, and Ryoji Takahashi.
\newblock {Gauge conditions for long term numerical black hole evolutions
  without excision}.
\newblock {\em Phys. Rev.}, D67:084023, 2003.

\bibitem{Brown:2008sb}
J.~David Brown, Peter Diener, Olivier Sarbach, Erik Schnetter, and Manuel
  Tiglio.
\newblock {Turduckening black holes: an analytical and computational study}.
\newblock {\em Phys. Rev. D}, 79:044023, 2009.

\bibitem{Baker:2002qf}
John~G. Baker, Manuela Campanelli, C.~O. Lousto, and R.~Takahashi.
\newblock {Modeling gravitational radiation from coalescing binary black
  holes}.
\newblock {\em Phys. Rev.}, D65:124012, 2002.

\bibitem{PhysRevD.83.044045}
Denis Pollney, Christian Reisswig, Erik Schnetter, Nils Dorband, and Peter
  Diener.
\newblock High accuracy binary black hole simulations with an extended wave
  zone.
\newblock {\em Phys. Rev. D}, 83:044045, Feb 2011.

\bibitem{Thornburg:1995cp}
Jonathan Thornburg.
\newblock {Finding apparent horizons in numerical relativity}.
\newblock {\em Phys. Rev. D}, 54:4899--4918, 1996.

\bibitem{Thornburg:2003sf}
Jonathan Thornburg.
\newblock {A Fast Apparent-Horizon Finder for 3-Dimensional Cartesian Grids in
  Numerical Relativity}.
\newblock {\em Class. Quant. Grav.}, 21:743--766, 2004.

\bibitem{Dreyer:2002mx}
Olaf Dreyer, Badri Krishnan, Deirdre Shoemaker, and Erik Schnetter.
\newblock {Introduction to Isolated Horizons in Numerical Relativity}.
\newblock {\em Phys. Rev.}, D67:024018, 2003.

\bibitem{Schnetter:2006yt}
Erik Schnetter, Badri Krishnan, and Florian Beyer.
\newblock {Introduction to dynamical horizons in numerical relativity}.
\newblock {\em Phys. Rev.}, D74:024028, 2006.

\bibitem{wardell_barry_2016_155394}
Barry Wardell, Ian Hinder, and Eloisa Bentivegna.
\newblock {Simulation of GW150914 binary black hole merger using the Einstein
  Toolkit}, September 2016.
\newblock \url{https://doi.org/10.5281/zenodo.155394}.

\bibitem{Healy:2014yta}
James Healy, Carlos~O. Lousto, and Yosef Zlochower.
\newblock {Remnant mass, spin, and recoil from spin aligned black-hole
  binaries}.
\newblock {\em Phys. Rev.}, D90(10):104004, 2014.

\bibitem{PhysRevD.95.024037}
James Healy and Carlos~O. Lousto.
\newblock Remnant of binary black-hole mergers: New simulations and peak
  luminosity studies.
\newblock {\em Phys. Rev. D}, 95:024037, Jan 2017.

\bibitem{RITcatalog:web}
{RIT Catalog for Numerical Simulations}.
\newblock \url{https://ccrg.rit.edu/~RITCatalog/}.

\bibitem{Ashtekar:2013qta}
Abhay Ashtekar, Miguel Campiglia, and Samir Shah.
\newblock {Dynamical Black Holes: Approach to the Final State}.
\newblock {\em Phys. Rev.}, D88(6):064045, 2013.

\bibitem{PhysRevLett.123.171102}
Daniel Pook-Kolb, Ofek Birnholtz, Badri Krishnan, and Erik Schnetter.
\newblock Interior of a binary black hole merger.
\newblock {\em Phys. Rev. Lett.}, 123:171102, Oct 2019.

\bibitem{PhysRevD.100.084044}
Daniel Pook-Kolb, Ofek Birnholtz, Badri Krishnan, and Erik Schnetter.
\newblock Self-intersecting marginally outer trapped surfaces.
\newblock {\em Phys. Rev. D}, 100:084044, Oct 2019.

\bibitem{PhysRevD.93.044006}
Sascha Husa, Sebastian Khan, Mark Hannam, Michael P\"urrer, Frank Ohme,
  Xisco~Jim\'enez Forteza, and Alejandro Boh\'e.
\newblock Frequency-domain gravitational waves from nonprecessing black-hole
  binaries. i. new numerical waveforms and anatomy of the signal.
\newblock {\em Phys. Rev. D}, 93:044006, Feb 2016.

\bibitem{PhysRevD.93.044007}
Sebastian Khan, Sascha Husa, Mark Hannam, Frank Ohme, Michael P\"urrer,
  Xisco~Jim\'enez Forteza, and Alejandro Boh\'e.
\newblock Frequency-domain gravitational waves from non precessing black-hole
  binaries. ii. a phenomenological model for the advanced detector era.
\newblock {\em Phys. Rev. D}, 93:044007, Feb 2016.

\bibitem{Booth:2006bn}
Ivan Booth and Stephen Fairhurst.
\newblock {Isolated, slowly evolving, and dynamical trapping horizons: geometry
  and mechanics from surface deformations}.
\newblock {\em Phys. Rev.}, D75:084019, 2007.

\bibitem{Datta:2019euh}
Sayak Datta and Sukanta Bose.
\newblock {Probing the nature of central objects in extreme-mass-ratio
  inspirals with gravitational waves}.
\newblock {\em Phys. Rev.}, D99(8):084001, 2019.

\bibitem{Datta:2019epe}
Sayak Datta, Richard Brito, Sukanta Bose, Paolo Pani, and Scott~A. Hughes.
\newblock {Tidal heating as a discriminator for horizons in extreme mass ratio
  inspirals}.
\newblock {\em Phys. Rev.}, D101(4):044004, 2020.

\end{thebibliography}

\end{document}